\documentclass[preprint,showpacs,amsmath,amssymb,eqsecnum]{revtex4}
\usepackage{dcolumn}
\usepackage{bm}
\usepackage{graphicx}

\begin{document}
\draft
\title{{Broken Flavor 2 \(\leftrightarrow\) 3 Symmetry and 
phenomenological approach for universal quark and lepton mass matrices}}

\author{Koichi MATSUDA}
\affiliation{%
Graduate school of Science, 
Osaka University, Toyonaka, Osaka 560-0043, Japan}
\author{Hiroyuki NISHIURA}
\affiliation{%
Faculty of Information Science and Technology, 
Osaka Institute of Technology, 
Hirakata, Osaka 573-0196, Japan}

\date{December 26, 2005}

\begin{abstract}
A phenomenological approach for the universal mass matrix model with a 
broken flavor $2 \leftrightarrow 3$ symmetry is explored by introducing 
the $2 \leftrightarrow 3$ antisymmetric parts of mass matrices for quarks 
and charged leptons. We present explicit texture components of the mass 
matrices, which are consistent with all the neutrino oscillation experiments 
and quark mixing data. 
The mass matrices have a common structure for quarks and leptons, 
while the large lepton mixings and the small quark mixings are derived with no fine tuning  
due to the difference of the phase factors. 
The model predicts a value $2.4\times10^{-3}$  
for the lepton mixing matrix element square $|U_{13}|^2$, and also  
$\langle m_\nu \rangle=(0.89-1.4)\times10^{-4}$ eV  for the averaged neutrino mass which appears in the neutrinoless 
double beta decay. 
\end{abstract}


\maketitle

\section{Introduction}
It has been established through the discovery of neutrino oscillation~\cite{skamioka} 
that neutrinos have finite masses and mix one another with near bimaximal lepton mixings $(\sin^2 2\theta_{12}\sim 1$, 
$\sin^2 2\theta_{23}\simeq 1)$ which are in contrast to small quark mixings.  
In order to explain the large lepton mixing and small quark mixing, 
mass matrix models with various structures have been investigated in the literature~\cite{fritzsch}--\cite{Kang}. 
For example, it is argued that the large lepton mixing can be explained 
by mass matrices with a flavor $2 \leftrightarrow 3$ symmetry~\cite{Fukuyama}-- \cite{Matsuda3}. 
We think that quarks and leptons should be unified. 
Therefore, it is interesting to investigate a possibility that 
all the mass matrices of the quarks and leptons have the same matrix form, 
which leads to the large lepton mixings and the small quark mixings.  
The mass matrix model with the universal form for quarks and leptons is also useful when it is embedded 
into a grand unified theory (GUT).
\par 
In this paper, 
we discuss a Hermit mass matrix model with a universal form given by 
\begin{equation}
M  =\left(
\begin{array}{lll}
\ 0 & \ ae^{-i\phi} & \ ae^{-i\phi^{\prime \prime}}  \\
\ ae^{i\phi} & \ b & \ ce^{-i\phi^\prime} \\
\ ae^{i\phi^{\prime \prime}} & \ ce^{i\phi^\prime} & \ b \\
\end{array}
\right) ,
\end{equation}
where $a$, $b$, and $c$ are real parameters and 
$\phi$, $\phi^\prime$, and $\phi^{\prime \prime}$ are phase parameters. 
It is important from a phenomenological point of view to parameterize the texture components of the mass matrix 
as the first step to make a GUT scenario. 
Assuming that neutrinos are the Majorana particles, we present the texture components of the universal mass matrices  
which will lead to the Cabibbo--Kobayashi--Maskawa (CKM) \cite{CKM} quark mixing
 and the Maki--Nakagawa--Sakata (MNS) \cite{MNS} lepton mixing 
which are consistent with the present experimental data.   
Here we explore a phenomenological mass matrix model base on the flavor $2 \leftrightarrow 3$ symmetry. 
Our mass matrices has a broken flavor $2 \leftrightarrow 3$ symmetry for quarks and charged leptons 
by introducing the $2 \leftrightarrow 3$ antisymmetric parts of their mass matrices. 
We assume that this broken flavor $2 \leftrightarrow 3$ symmetry is due to the {\bf{120}} Higgs scalar in the SO(10) GUT model, 
while mass matrices contributed from {\bf{10}} and {\bf{126}} Higgs scalars are $2 \leftrightarrow 3$ symmetric.
\par
This article is organized as follows. 
In Sec.~II, our mass matrix model is presented. 
In Sec.~III, we discuss the diagonalization of mass matrix of our model. 
The analytical expressions of the quark and lepton mixings of the model are given in Sec.~IV. 
Sec.~V  is devoted to a summary.

\section{Mass matrix Model}
\par
In this paper, we propose the following mass matrices: 
\begin{eqnarray}
M_u  &=& \left(
\begin{array}{lll}
\ 0 & \ \frac{1}{\sqrt{2}}e^{-i\phi_u}A_u & \ \frac{1}{\sqrt{2}}e^{-i\phi_u}A_u \\
\ \frac{1}{\sqrt{2}}e^{i\phi_u}A_u & \ \frac{B_u+D_u}{2} & \ \frac{B_u-D_u}{2} \\
\ \frac{1}{\sqrt{2}}e^{i\phi_u}A_u  & \ \frac{B_u-D_u}{2} & \ \frac{B_u+D_u}{2}  \\
\end{array}
\right) +
\left(
\begin{array}{lll}
\ 0 & \ 0 & 0 \\
\ 0 & \ 0 & iC_u \\
\ 0  & \ -iC_u & 0  \\
\end{array}
\right) , \\
M_d  &=& \left(
\begin{array}{lll}
\ 0 & \ \frac{1}{\sqrt{2}}e^{-i\phi_d}A_d & \ \frac{1}{\sqrt{2}}e^{-i\phi_d}A_d \\
\ \frac{1}{\sqrt{2}}e^{i\phi_d}A_d & \ \frac{B_d+D_d}{2} & \ \frac{B_d-D_d}{2}  \\
\ \frac{1}{\sqrt{2}}e^{i\phi_d}A_d  & \ \frac{B_d-D_d}{2} & \ \frac{B_d+D_d}{2}  \\
\end{array}
\right) +
\left(
\begin{array}{lll}
\ 0 & \ 0 & 0 \\
\ 0 & \ 0 & iC_d \\
\ 0  & \ -iC_d & 0  \\
\end{array}
\right) , \\
M_e  &=& \left(
\begin{array}{lll}
\ 0 & \ \frac{1}{2}A_e & \ \frac{1}{2}A_e  \\
\ \frac{1}{2}A_e & \ \frac{B_e+D_e}{2} & \ -C_e \\
\ \frac{1}{2}A_e & \ -C_e\ & \ \frac{B_e+D_e}{2} \\
\end{array}
\right) +
\left(
\begin{array}{lll}
\ 0 & \ -i\frac{1}{2}A_e & \ i\frac{1}{2}A_e  \\
\ i\frac{1}{2}A_e & \ \ \ 0 & \ i\frac{B_e-D_e}{2} \\
\ -i\frac{1}{2}A_e & \ -i\frac{B_e-D_e}{2} & \ \ \ 0 \\
\end{array}
\right) ,  \\
M_\nu   &=& \left(
\begin{array}{lll}
\ 0 & \ \frac{1}{\sqrt{2}}A_\nu & \ \frac{1}{\sqrt{2}}A_\nu \\
\ \frac{1}{\sqrt{2}}A_\nu & \ \frac{B_\nu-D_\nu}{2} & \ \frac{B_\nu+D_\nu}{2} \\
\ \frac{1}{\sqrt{2}}A_\nu  & \ \frac{B_\nu+D_\nu}{2} & \ \frac{B_\nu-D_\nu}{2} \\
\end{array}
\right) ,\\
M_D   &=& \left(
\begin{array}{lll}
\ 0 & \ \frac{1}{\sqrt{2}}A_D & \ \frac{1}{\sqrt{2}}A_D \\
\ \frac{1}{\sqrt{2}}A_D & \ \frac{B_D+D_D}{2} & \ \frac{B_D-D_D}{2} \\
\ \frac{1}{\sqrt{2}}A_D  & \ \frac{B_D-D_D}{2} & \ \frac{B_D+D_D}{2} \\
\end{array}
\right) , \\
M_R  &=& \left(
\begin{array}{lll}
\ 0 & \ \frac{1}{\sqrt{2}}A_R & \ \frac{1}{\sqrt{2}}A_R \\
\ \frac{1}{\sqrt{2}}A_R & \ \frac{B_R+D_R}{2} & \ \frac{B_R-D_R}{2} \\
\ \frac{1}{\sqrt{2}}A_R  & \ \frac{B_R-D_R}{2} & \ \frac{B_R+D_R}{2} \\
\end{array}
\right) ,  
\end{eqnarray}
where \(M_u\), \(M_d\), \(M_e\), and \(M_\nu \) are mass matrices 
for up quarks (\(u,c,t\)), down quarks (\(d,s,b\)),  charged leptons (\(e,\mu,\tau \)), 
and  neutrinos (\(\nu_e,\nu_{\mu},\nu_{\tau} \)), respectively.  
The mass matrices \(M_D\) and \(M_R\) are, respectively, 
the Dirac and the right-handed Majorana type neutrino mass matrices, 
from which with the seesaw mechanism \cite{Yanagida} we derive \(M_\nu \).
Here $A_f$, $B_f$, $C_f$, and $D_f$ are real parameters and 
$\phi_f$ and $\phi_f^\prime$ are phase parameters with $f=u, d, e$, and $\nu$. 
\par
Let us mention a particular feature of these mass matrices with respect to the flavor $2 \leftrightarrow 3$ symmetry.
We assume that the neutrino mass matrix has only $2 \leftrightarrow 3$ symmetric part. 
In the mass matrices for quarks and charged leptons, 
the $2 \leftrightarrow 3$ anti-symmetric terms (the second terms) are added as broken  $2 \leftrightarrow 3$ symmetric parts, 
in addition to the $2 \leftrightarrow 3$ symmetric terms (the first terms). 
This structure is motivated by the SO(10) GUT model in which  {\bf{10}}, {\bf{120}}, and {\bf{126}} Higgs scalars contribute to the fermion mass matrices,  
together with the following assumptions: 
 (i) The contribution from the {\bf{120}} Higgs scalar is $2 \leftrightarrow 3$ anti-symmetric, 
while those from {\bf{10}} and {\bf{126}} Higgs scalars are $2 \leftrightarrow 3$ symmetric for quarks and charged leptons.
(ii) There exists the contribution to the Dirac type neutrino mass matrix $M_D$ from only the {\bf{10}} and {\bf{126}} Higgs scalars.
and (iii) The texture components of the broken $2 \leftrightarrow 3$ symmetric parts are assumed 
to have different form between quarks and charged leptons, 
which derives a difference between the small quark mixing and the large lepton mixing. 
Namely, we assume that the mass matrices $M_u$ and $M_d$ are superpositions of the common real symmetric matrices ${\cal{S}}$ and  ${\cal{S^\prime}}$ 
and pure imaginary anti-symmetric one ${\cal{A}}$ and that $M_e$, $M_D$, and $M_R$ consist of the common real symmetric matrices $\cal{S^{\prime \prime}}$ and $\cal{S^{\prime \prime \prime}}$ 
and pure imaginary anti-symmetric one $\cal{A^\prime}$,  
as follows. 
\begin{eqnarray}
M_u  &=&
\alpha_u {\cal{S}}+
\beta_u {\cal{S^{\prime}}} +
\gamma_u {\cal{A}},  \\
M_d  &=&
\alpha_d {\cal{S}}+
\beta_d {\cal{S^{\prime}}} +
\gamma_d {\cal{A}},  \\
M_e &=&
\alpha_e {\cal{S^ {\prime \prime}}}+\beta_e {\cal{S^{\prime \prime \prime}}} +
\gamma_e {\cal{A^\prime}},\\
M_D &=&
\alpha_D {\cal{S^{\prime \prime}}}+\beta_D {\cal{S^{\prime \prime \prime}}},  \\
M_R &=& \beta_R {\cal{S^{\prime \prime \prime}}}, \\  
M_\nu &=& -M_D^T M_R^{-1}M_D, 
\end{eqnarray}
where the matrices ${\cal{S}}$, ${\cal{S^\prime}}$, $\cal{S^{\prime \prime}}$ and $\cal{S^{\prime \prime \prime}}$ 
are $2 \leftrightarrow 3$ symmetric too, 
and ${\cal{A}}$ and $\cal{A^\prime}$ are $2 \leftrightarrow 3$ anti-symmetric too. 
Here $\alpha_i$, $\beta_i$, $\gamma_i$ ($i=u,d,e$), $\alpha_D$, $\beta_D$, and $\beta_R$ are real coefficient parameters.
Note that the $2 \leftrightarrow 3$ symmetry of the model is broken through only ${\cal{A}}$ in the quark sector and ${\cal{A}}^\prime$ in the lepton sector.
\par
Some semi-empirical approaches for mass matrices with the similar
structure to the above Eqs.~(2.7) - (2.12) have been proposed in the
literature. For example, Gronau, Johnson, and Schechter~\cite{Gronau}
have discussed a model which consists of combining the Fritzch~\cite{fritzsch}
and Stech~\cite{Stech} ansatz for quarks.  They use the combination of
symmetric mass matrix with antisymmetric one, although they don't use
the $2 \leftrightarrow 3$ symmetry.  An extension to leptons based on an
SO(10) GUT model has been investigated with use of the type I and type
II seesaw mechanism for neutrino masses~\cite{Bottino,Johnson}. 
In the present paper, we use the $2 \leftrightarrow 3$ symmetry
for a common origin of the small quark and the large lepton mixings.
This is the large difference between our model and the other 
2\(\leftrightarrow\)3 symmetry models \cite{Fukuyama}-\cite{Mohapatra2}.
\par
The mass matrix $M_f$ ($f=u, d, e$, and $\nu$) given in Eqs.~(2.1)--(2.4) has common structure 
when it is expressed with a unitary matrix $Q_{f}$ as follows:
\begin{eqnarray}
M_f &=& 
Q_{f} 
\widehat{M_f}
Q_{f}^\dagger , \quad \mbox{for $f=u$, $d$, and $e$} \nonumber\\ 
M_f &=& 
Q_{f} 
\widehat{M_f}
Q_{f}^T , \quad \mbox{for $f=\nu$}\label{M}
\end{eqnarray}
where $\widehat{M_f}$ ($f=u, d, e$, and $\nu$) is one of the seesaw-invariant type of mass matrix defined by~\cite{Nishiura2}
\begin{equation}
\widehat{M_f} = 
\left(
\begin{array}{lll}
\ 0 & \ A_f & \ 0 \\
\ A_f & \ B_f & \ C_f \\
\ 0 & \ C_f & \ D_f \\
\end{array}
\right) .
\ \ \label{M-hat}
\end{equation}
Here the unitary matrices $Q_{f}$ are given by 
\begin{eqnarray}
Q_{u}&=&
\left(
\begin{array}{lll}
\ 1 & \ \ \ \ \  0 \  & \ \ \ \ \ \   0 \  \\
\ 0 & \ \ \ \frac{1}{\sqrt{2}}e^{i\phi_u} & \ \ \ \frac{1}{\sqrt{2}} i e^{i\phi_u} \  \\
\ 0 & \ \ \ \frac{1}{\sqrt{2}}e^{i\phi_u}  & \ -\frac{1}{\sqrt{2}}  i e^{i\phi_u} \ \\
\end{array}
\right), \\ 
Q_{d}&=&
\left(
\begin{array}{lll}
\ 1 & \ \ \ \ \  0 \  & \ \ \ \ \ \   0 \  \\
\ 0 & \ \ \ \frac{1}{\sqrt{2}}e^{i\phi_d} & \ \ \ \frac{1}{\sqrt{2}} i e^{i\phi_d} \  \\
\ 0 & \ \ \ \frac{1}{\sqrt{2}}e^{i\phi_d}  & \ -\frac{1}{\sqrt{2}}  i e^{i\phi_d} \ \\
\end{array}
\right), \\ 
Q_{e}&=&
\left(
\begin{array}{lll}
\ 1 & \ \  \ \ \ 0 \  & \ \ \ \ \   0 \  \\
\ 0 & \ \ \ \frac{1}{\sqrt{2}}e^{i\frac{\pi}{4}} & \ \ \ \frac{1}{\sqrt{2}} i e^{i\frac{\pi}{4}} \  \\
\ 0 & \ \ \ \frac{1}{\sqrt{2}}e^{-i\frac{\pi}{4}}  & \ -\frac{1}{\sqrt{2}}  i e^{-i\frac{\pi}{4}} \ \\
\end{array}
\right), \\ 
Q_{\nu}&=&
\left(
\begin{array}{lll}
\ 1 & \ \ \ \ 0 \  & \ \ \ \  0 \  \\
\ 0 & \ \ \ \frac{1}{\sqrt{2}} & \ \ \  \frac{1}{\sqrt{2}} i  \  \\
\ 0 & \ \ \  \frac{1}{\sqrt{2}}  & \ -\frac{1}{\sqrt{2}}  i  \ \\
\end{array}
\right) . 
\end{eqnarray}
Note that the structure of $Q_f$ mentioned above is the same for all the quarks and leptons 
except for the phase factors in it. 
It should be also noted that the Eq.~(\ref{M}) implies that the mass matrix $M_f$ is transformed to  
$\widehat{M_f}$ by using a rebasing of the quark and lepton fields respectively.
\section{Diagonalization of Mass matrix}
\par
We now discuss a diagonalization of the mass matrix $M_f$ given in Eq.~(\ref{M}). 
First let us discuss the diagonalization of the mass matrix $\widehat{M_f}$ given in Eq.~(\ref{M-hat}), which appears as a part of $M_f$.
This $\widehat{M_f}$ is diagonalized  by an orthogonal matrix $O_f$ as discussed in Refs.~\cite{Koide} and \cite{Matsuda}; 
\begin{equation}
O_f^T\left(
\begin{array}{lll}
\ 0 & \ A_f & \ 0 \\
\ A_f & \ B_f & \ C_f \\
\ 0 & \ C_f & \ D_f \\
\end{array}
\right)
O_f
=\left(
\begin{array}{lll}
-m_{1f} & \  & \  \\
\  & m_{2f} & \  \\
\  & \  & m_{3f} \\
\end{array}
\right).
\end{equation}
Here $m_{1f}$, $m_{2f}$, and $m_{3f}$ are eigenvalues of $M_f$.
Explicit expressions of the orthogonal matrix $O_f$, and components $A_f$, $B_f$, $C_f$, and $D_f$ 
in terms of $m_{1f}$, $m_{2f}$, and $m_{3f}$ are presented in Appendix A.
Namely,  the mass matrix $M_f$ is diagonalized as
\begin{eqnarray}
U_{Lf}^\dagger M_f  U_{Lf} 
&=&
\left(
\begin{array}{lll}
-m_{1f} & \  & \  \\
\  & m_{2f} & \  \\
\  & \  & m_{3f} \\
\end{array}
\right)\quad \mbox{for $f=u$, $d$, and $e$} \ ,\ \\
U_{Lf}^\dagger M_f  U_{Lf}^{*} 
&=&
\left(
\begin{array}{lll}
-m_{1f} & \  & \  \\
\  & m_{2f} & \  \\
\  & \  & m_{3f} \\
\end{array}
\right) \quad \mbox{for $f=\nu$}. 
\end{eqnarray}
where the unitary matrix $U_{Lf}$ is given by\\
\begin{equation}
U_{Lf}=Q_{f} O_f .
\end{equation}
Here we list the expressions for $O_f$ and $Q_f$ in order:
\begin{eqnarray}
O_f
&\simeq&
\left(
        \begin{array}{ccc}
        1& \sqrt{\frac{m_{1f}}{m_{2f}}}&
         \sqrt{\frac{m_{1f}m_{2f}^2}{m_{3f}^3}}\\
        -\sqrt{\frac{m_{1f}}{m_{2f}}}
        & 1 & \sqrt{\frac{m_{1f}}{m_{3f}}}\\
        \sqrt{\frac{{m_{1f}^2}}{m_{2f}m_{3f}}}
        & -\sqrt{\frac{m_{1f}}{m_{3f}}} & 1
        \end{array}
\right) \quad \mbox{for $f=u$, $d$, and $e$}\ ,\\
O_\nu
&=&
\left(
        \begin{array}{ccc}
        \sqrt{\frac{m_2}{m_2+m_1}}& \sqrt{\frac{m_1}{m_2+m_1}}&
         0\\
        -\sqrt{\frac{m_1}{m_2+m_1}}
        & \sqrt{\frac{m_2}{m_2+m_1}} & 0\\
        0
        & 0& 1
        \end{array}
\right) ,
\end{eqnarray}
and 
\begin{eqnarray}
Q_{f}&=&
\left(
\begin{array}{lll}
\ 1 & \ \ \ \ \  0 \  & \ \ \ \ \ \   0 \  \\
\ 0 & \ \ \ \frac{1}{\sqrt{2}}e^{i\phi_f} & \ \ \ \frac{1}{\sqrt{2}} i e^{i\phi_f} \  \\
\ 0 & \ \ \ \frac{1}{\sqrt{2}}e^{i\phi_f}  & \ -\frac{1}{\sqrt{2}}  i e^{i\phi_f} \ \\
\end{array}
\right) \quad \mbox{for $f=u$ and $d$ } ,\label{Q-ud}\\
Q_{e}&=&
\left(
\begin{array}{lll}
\ 1 & \ \  \ \ \ 0 \  & \ \ \ \ \   0 \  \\
\ 0 & \ \ \ \frac{1}{\sqrt{2}}e^{i\frac{\pi}{4}} & \ \ \ \frac{1}{\sqrt{2}} i e^{i\frac{\pi}{4}} \  \\
\ 0 & \ \ \ \frac{1}{\sqrt{2}}e^{-i\frac{\pi}{4}}  & \ -\frac{1}{\sqrt{2}}  i e^{-i\frac{\pi}{4}} \ \\
\end{array}
\right) ,  \label{Q-e}\\
Q_{\nu}&=&
\left(
\begin{array}{lll}
\ 1 & \ \ \ \ 0 \  & \ \ \ \  0 \  \\
\ 0 & \ \ \ \frac{1}{\sqrt{2}} & \ \ \  \frac{1}{\sqrt{2}} i  \  \\
\ 0 & \ \ \  \frac{1}{\sqrt{2}}  & \ -\frac{1}{\sqrt{2}}  i  \ \\
\end{array}
\right) . \label{Q-nu}
\end{eqnarray}
Here, $m_{iu}$, $m_{id}$, $m_{ie}$, and $m_{i\nu}\ (i=1,2,3) $ are, respectively, 
the masses of up quarks, down quarks, charged leptons, and neutrinos, 
which we shall denote as $(m_u, m_c, m_t)$, $(m_d, m_s, m_b)$, $(m_e, m_\mu, m_\tau)$ and $(m_1, m_2, m_3)$.
\par
Furthermore, the neutrino mass matrix is diagonalized as 
\begin{equation}
U_{L\nu}^{\prime \dagger} M_f  U_{L\nu}^{\prime *} =
\left(
\begin{array}{lll}
m_{1} & \  & \  \\
\  & m_{2} & \  \\
\  & \  & m_{3} \\
\end{array}
\right), 
\end{equation}
where the unitary matrix $U_{L\nu}^\prime$ is given by
\begin{equation}
U_{L\nu}^\prime=U_{L\nu}P_\nu=Q_{f} O_fP_\nu .
\end{equation}
Here, in order to make the neutrino masses to be real positive,  
we introduced a diagonal phase matrix $P_\nu$ defined by 
\begin{equation}
P_\nu=\mbox{diag}(i,1,1).
\end{equation}

\section{CKM quark and MNS lepton mixing matrices}
\par
Next we discuss the CKM quark mixing matrix $V$ and the MNS lepton mixing matrix $U$ 
of the model, which are given by 
\begin{eqnarray}
V&=&U^\dagger_{Lu}U_{Ld}=O^{T}_uQ^\dagger_{u} Q_{d}O_d , \\
U&=&U^\dagger_{Le}U_{L\nu}^\prime=O^{T}_eQ^\dagger_{e} Q_{\nu}O_\nu P_\nu.
\end{eqnarray}
From Eqs.(\ref{Q-ud}) -- (\ref{Q-nu}), we obtain
\begin{eqnarray}
Q^\dagger_{u} Q_{d} &=&
\left(
\begin{array}{lll}
\ 1 & \ \ \ 0 \  & \ 0 \  \\
\ 0 & \ e^{i(\phi_d-\phi_u)} & \ 0 \  \\
\ 0 & \ \ \ 0 & e^{i(\phi_d-\phi_u)}\ \\
\end{array}
\right), \label{QQ-q}\\ 
Q^\dagger_{e} Q_{\nu} &=&
\left(
\begin{array}{lll}
\ 1 & \ \ \ \ 0 \  & \ \ \ 0 \  \\
\ 0 & \ \ \ \frac{1}{\sqrt{2}} & \ \  \frac{1}{\sqrt{2}}   \  \\
\ 0 & \ -\frac{1}{\sqrt{2}}  & \ \ \frac{1}{\sqrt{2}}   \ \\
\end{array}
\right). \label{QQ-l}
\end{eqnarray}
It should be noted that $Q^\dagger_{e} Q_{\nu}$ 
takes quite different structure from that of $Q^\dagger_{u} Q_{d}$ in our model.
Namely, $Q^\dagger_{u} Q_{d}$ is a diagonal phase matrix, 
while $Q^\dagger_{e} Q_{\nu}$ represents a mixing matrix with a maximal lepton mixing 
between the second and third generations. Therefore,  
the large lepton mixing is realized with no fine tuning in our model.
\par
Let us discuss the quark and lepton mixing matrices in detail.
\subsection{CKM quark mixing matrix} 
We obtain the CKM quark mixing matrix $V$ as follows:
\begin{eqnarray}
V&=&O^{T}_uQ^\dagger_{u} Q_{d}O_d\\
&=&\left(
        \begin{array}{ccc}
        1& \sqrt{\frac{m_{u}}{m_{c}}}&
         \sqrt{\frac{m_{u}m_{c}^2}{m_{t}^3}}\\
        -\sqrt{\frac{m_{u}}{m_{c}}}
        & 1 & \sqrt{\frac{m_{u}}{m_{t}}}\\
        \sqrt{\frac{{m_{u}^2}}{m_{c}m_{t}}}
        & -\sqrt{\frac{m_{u}}{m_{t}}} & 1
        \end{array}
\right)^{T}
\left(
\begin{array}{lll}
\ 1 & \ \ \ 0 \  & \ 0 \  \\
\ 0 & \ e^{i(\phi_d-\phi_u)} & \ 0 \  \\
\ 0 & \ \ \ 0 & e^{i(\phi_d-\phi_u)}\ \\
\end{array}
\right) \nonumber\\
&&\times\left(
        \begin{array}{ccc}
        1& \sqrt{\frac{m_{d}}{m_{s}}}&
         \sqrt{\frac{m_{d}m_{s}^2}{m_{b}^3}}\\
        -\sqrt{\frac{m_{d}}{m_{s}}}
        & 1 & \sqrt{\frac{m_{d}}{m_{b}}}\\
        \sqrt{\frac{{m_{d}^2}}{m_{s}m_{b}}}
        & -\sqrt{\frac{m_{d}}{m_{b}}} & 1
        \end{array}
\right). \label{our_ckm}
\end{eqnarray}
The explicit magnitudes of $(i,j)$ elements of $V$ are obtained as
\begin{eqnarray}
|V_{12}|&\simeq&\left|\sqrt{\frac{m_{d}}{m_{s}}}-\sqrt{\frac{m_{u}}{m_{c}}}e^{i(\phi_d-\phi_u)}\right|=|0.224-0.06e^{i(\phi_d-\phi_u)}|,\label{V12}\\
|V_{23}|&\simeq&\left|\sqrt{\frac{m_{d}}{m_{b}}}-\sqrt{\frac{m_{u}}{m_{t}}}\right|=0.0336,\label{V23}\\
|V_{13}|&\simeq&\left|\sqrt{\frac{m_{d}m_{s}^2}{m_{b}^3}}-\sqrt{\frac{m_{u}m_{d}}{m_{c}m_{b}}}e^{i(\phi_d-\phi_u)}\right| \nonumber \label{V13}\\
&=&|0.00022-0.0021e^{i(\phi_d-\phi_u)}|.
\end{eqnarray}
Here we have used the following numerical values for the quark masses estimated at the unification scale \(\mu=M_X\), which are  presented in Appendix A.
\begin{equation}
\begin{array}{ll}
m_u(M_X)=1.04^{+0.19}_{-0.20}\, \mbox{MeV},& 
m_d(M_X)=1.33^{+0.17}_{-0.19}\, \mbox{MeV}, \\
m_c(M_X)=302^{+25}_{-27}\, \mbox{MeV}, &
m_s(M_X)=26.5^{+3.3}_{-3.7}\, \mbox{MeV}, \\ 
m_t(M_X)=129^{+196}_{-40}\,  \mbox{GeV}, &
m_b(M_X)=1.00\pm0.04\, \mbox{GeV}. 
\end{array}
\label{eq123104}
\end{equation}
\par
By using the rephasing of the up and down quarks, 
Eq.~(\ref{our_ckm}) is changed to the standard representation of the CKM quark mixing matrix, 
\begin{eqnarray}
V_{\rm std} &=& \mbox{diag}(e^{i\zeta_1^u},e^{i\zeta_2^u},e^{i\zeta_2^u})  \ V \ 
\mbox{diag}(e^{i\zeta_1^d},e^{i\zeta_2^d},e^{i\zeta_2^d}) \nonumber \\
&=&
\left(
\begin{array}{ccc}
c_{13}c_{12} & c_{13}s_{12} & s_{13}e^{-i\delta} \\
-c_{23}s_{12}-s_{23}c_{12}s_{13} e^{i\delta}
&c_{23}c_{12}-s_{23}s_{12}s_{13} e^{i\delta} 
&s_{23}c_{13} \\
s_{23}s_{12}-c_{23}c_{12}s_{13} e^{i\delta}
 & -s_{23}c_{12}-c_{23}s_{12}s_{13} e^{i\delta} 
& c_{23}c_{13} \\
\end{array}
\right) \ .
\label{stdrep}
\end{eqnarray}
Here \(\zeta_i^q\) comes from the rephasing in the quark fields 
to make the choice of phase convention.
The $CP$ violating phase \(\delta\) in Eq.~(\ref{stdrep}) 
is predicted with the expression of $V$ in Eq.~(\ref{our_ckm}) as 
\begin{equation}
\delta =
\mbox{arg}\left[
\left(\frac{V_{us} V_{cs}^{*}}{V_{ub} V_{cb}^{*}}\right) + 
\frac{|V_{us}|^2}{1-|V_{ub}|^2}
\right]\simeq \phi_u-\phi_d +\pi.\label{delta}
\end{equation}
\par
The predicted values of $|V_{12}|$, $|V_{23}|$, $|V_{13}|$, and $\delta$ 
are functions of a free parameter $\phi_u-\phi_d$ as shown in Eqs.~(\ref{V12})--(\ref{V13}) and (\ref{delta}).
They are roughly consistent with the following numerical values at \(\mu=M_X\), 
which are estimated from the experimental data observed at electroweak scale \(\mu=M_Z\) 
by using the renormalization group equation and presented in Appendix B:
\begin{eqnarray}
|V_{12}^0|&=&0.2226-0.2259,\nonumber\\
|V_{23}^0|&=&0.0295-0.0387,\\
|V_{13}^0|&=&0.0024-0.0038,\\
\delta^0 &=& 46^\circ   - 74^\circ . \nonumber
\end{eqnarray}
\subsection{MNS lepton mixing matrix} 
We obtain the MNS lepton mixing matrix \(U\) as follows: 
\begin{eqnarray}
U&=&O^{T}_eQ^\dagger_{e} Q_{\nu}O_\nu P_\nu\\
&=&
\left(
        \begin{array}{ccc}
        1& \sqrt{\frac{m_{e}}{m_{\mu}}}&
         \sqrt{\frac{m_{e}m_{\mu}^2}{m_{\tau}^3}}\\
        -\sqrt{\frac{m_{e}}{m_{\mu}}}
        & 1 & \sqrt{\frac{m_{e}}{m_{\tau}}}\\
        \sqrt{\frac{{m_{e}^2}}{m_{\mu}m_{\tau}}}
        & -\sqrt{\frac{m_{e}}{m_{\tau}}} & 1
        \end{array}
\right)^{T}  \nonumber\\ 
&&
\times
\left(
\begin{array}{lll}
\ 1 & \ \ \ \ 0 \  & \ \ \ 0 \  \\
\ 0 & \ \ \ \frac{1}{\sqrt{2}} & \ \  \frac{1}{\sqrt{2}}   \  \\
\ 0 & \ -\frac{1}{\sqrt{2}}  & \ \ \frac{1}{\sqrt{2}}   \ \\
\end{array}
\right)
\left(
        \begin{array}{ccc}
        \sqrt{\frac{m_2}{m_2+m_1}}& \sqrt{\frac{m_1}{m_2+m_1}}&
         0\\
        -\sqrt{\frac{m_1}{m_2+m_1}}
        & \sqrt{\frac{m_2}{m_2+m_1}} & 0\\
        0
        & 0& 1
        \end{array}
\right)P_\nu \nonumber \\
&\simeq&
\left(
\begin{array}{lll}
\ \ \ \ c_1i & \ \ \ \ s_1 \  & \  -\frac{1}{\sqrt{2}}\sqrt{\frac{m_{e}}{m_{\mu}}} \  \\
\ -\frac{1}{\sqrt{2}}s_1i & \ \ \ \frac{1}{\sqrt{2}}c_1 & \ \  \frac{1}{\sqrt{2}}   \  \\
\ \ \ \frac{1}{\sqrt{2}}s_1i & \ -\frac{1}{\sqrt{2}}c_1  & \ \ \frac{1}{\sqrt{2}}   \ \\
\end{array}
\right), \label{ourMNS} 
\end{eqnarray}
with 
\begin{equation}
s_1\equiv\sqrt{\frac{m_1}{m_2+m_1}}, \quad c_1\equiv\sqrt{\frac{m_2}{m_2+m_1}}.
\end{equation}
The explicit magnitudes of $(i,j)$ elements of $U$ are
\begin{equation}
|U|
\simeq
\left(
\begin{array}{ccc}
\sqrt{\frac{m_2}{m_2+m_1}} \quad & \sqrt{\frac{m_1}{m_2+m_1}}
\quad & \frac{1}{\sqrt{2}} \sqrt{\frac{m_e}{m_\mu}}\\
\frac{1}{\sqrt{2}}\sqrt{\frac{m_1}{m_2+m_1}} \quad & \frac{1}{\sqrt{2}}\sqrt{\frac{m_2}{m_2+m_1}} \quad & \frac{1}{\sqrt{2}}\\
\frac{1}{\sqrt{2}}\sqrt{\frac{m_1}{m_2+m_1}} \quad & \frac{1}{\sqrt{2}}\sqrt{\frac{m_2}{m_2+m_1}} 
\quad & \frac{1}{\sqrt{2}} \\
\end{array}
\right),
\end{equation}
Therefore, we obtain 
\begin{eqnarray}
\tan^2\theta_{\mbox{{\tiny solar}}}& =&\frac{|U_{12}|^2}{|U_{11}|^2}\simeq \frac{m_1}{m_2}\ ,\label{eq30300}\\
\sin^2 2\theta_{\mbox{{\tiny atm}}}& =&4|U_{23}|^2|U_{33}|^2\simeq 1\ ,
\label{eq30310} \\
|U_{13}|^2 &\simeq& \frac{m_e}{2m_\mu}.   
\end{eqnarray}
\par
In the following  discussions we consider the normal mass hierarchy 
$m_1 < m_2 \ll m_3$ for the neutrino masses.
Then the evolution effects which only give negligibly small correction effects can be ignored. 
Scenarios in which the neutrino masses have the quasi degenerate 
or the inverse hierarchy  
will be denied from Eqs.~(\ref{eq30300}) and (\ref{eq20501}). 
\par
It can be seen from Eq.~(\ref{ourMNS}) that the large lepton mixing angle between the second and third generation is well realized with no fine tuning in the model.
It should be noted that the present model leads to the same results 
for $\theta_{\mbox{{\tiny solar}}}$ and $\theta_{\mbox{{\tiny atm}}}$ as the model in Ref~\cite{Matsuda2}, 
while a different feature for $|U_{13}|^2$ is derived.
\par
On the other hand, we have~\cite{Garcia} a experimental bound for $|U_{13}|_{\mbox{\tiny exp}}^2$
from the CHOOZ~\cite{chooz}, solar~\cite{sno}, and atmospheric neutrino 
experiments~\cite{skamioka}. 
From the global analysis of the SNO solar neutrino experiment~\cite{sno,Garcia}, 
we have $\Delta m_{12}^2$ and $\tan^2 \theta_{12}$ for the large mixing angle (LMA) Mikheyev-Smirnov-Wolfenstein (MSW) solution.
From the atmospheric neutrino experiment~\cite{skamioka,Garcia} , 
we also have $\Delta m_{23}^2$ and $\tan^2 \theta_{23}$. 
These experimental data with $3\sigma$ range are given by 
\begin{eqnarray}
& &|U_{13}|_{\mbox{\tiny exp}}^2 <  0.054 \ \label{mat20820} \ ,\\
& &\Delta m_{12}^2=m_2^2-m_1^2= \Delta m_{\mbox{{\tiny sol}}}^2
=(5.2-9.8) \times 10^{-5}\, \mbox{eV}^2, \label{mat20830} \\
& &\tan^2 \theta_{12}=\tan^2 \theta_{\mbox{{\tiny sol}}}=0.29-0.64 \ ,\label{eq20501}\\
& &\Delta m_{23}^2=m_3^2-m_2^2 \simeq \Delta m_{\mbox{{\tiny atm}}}^2
= (1.4-3.4) \times 10^{-3}\, \mbox{eV}^2, \label{mat20831}\\ 
& &\tan^2 \theta_{23} \simeq \tan^2 \theta_{\mbox{{\tiny atm}}}=0.49-2.2 \ . \label{mat208302}
\end{eqnarray}
Hereafter, for simplicity, we take  $\tan^2 \theta_{\mbox{{\tiny atm}}} \simeq 1$.
Thus, by combining the present model with the mixing angle \(\theta_{\mbox{{\tiny sol}}}\),
we have 
\begin{equation}
\frac{m_1}{m_2} \simeq \tan^2\theta_{\mbox{{\tiny sol}}}=0.29 - 0.64. 
\label{ratio}
\end{equation}
Therefore we predict the neutrino masses as follows.
\begin{eqnarray}
m_1^2 & = & (0.48-6.8) \times 10^{-5} \  {\rm eV^2} \ ,\nonumber \\
m_2^2 & = & (5.7-16.6) \times 10^{-5} \  {\rm eV^2} \ ,  \label{neu-mass}\\
m_3^2 & = & (1.4-3.4) \times 10^{-3} \  {\rm eV^2} \ .\nonumber
\end{eqnarray}
Let us mention other predictions in our model. 
Our model imposes a restriction on \(|U_{13}|\) as
\begin{equation}
|U_{13}|^2 \simeq  \frac{m_e}{2m_{\mu}}=2.4\times10^{-3}  . \label{mat20870} 
\end{equation}
Here we have used the running charged lepton masses at the unification scale \(\mu=\Lambda_X\)
\cite{Fusaoka}: $m_e(\Lambda_X)=0.325\ \mbox{MeV}$, 
$m_\mu(\Lambda_X)=68.6\ \mbox{MeV}$, 
and $m_\tau(\Lambda_X)=1171.4 \pm 0.2\ \mbox{MeV}$.
The value in Eq.~(\ref{mat20870}) is consistent with the present experimental constraints 
Eq.~(\ref{mat20820}). 

\par
Next let us discuss the CP-violation phases in the lepton mixing matrix.
The Majorana neutrino fields do not have the freedom of rephasing 
invariance, so that we can use only the rephasing freedom of $M_e$ 
to transform Eq.~(\ref{ourMNS}) to the standard form
\begin{eqnarray}
& &U_{\rm std} 
= \mbox{diag}(e^{i\alpha_1^e},e^{i\alpha_2^e},e^{i\alpha_2^e}) 
\ U  \nonumber \\ 
& &= \left(
\begin{array}{ccc}
c_{\nu13}c_{\nu12} & c_{\nu13}s_{\nu12}e^{i\beta} & 
s_{\nu13}e^{i(\gamma-\delta_{\nu})} \\
(-c_{\nu23}s_{\nu12}-s_{\nu23}c_{\nu23}s_{\nu13} e^{i\delta_{\nu}})e^{-i\beta}
&c_{\nu23}c_{\nu12}-s_{\nu23}s_{\nu12}s_{\nu13} e^{i\delta_{\nu}} 
&s_{\nu23}c_{\nu13}e^{i(\gamma-\beta)} \\
(s_{\nu23}s_{\nu12}-c_{\nu23}c_{\nu12}s_{\nu13} e^{i\delta_{\nu}})e^{-i\gamma}
 & (-s_{\nu23}c_{\nu12}-c_{\nu23}s_{\nu12}s_{\nu13} 
e^{i\delta_{\nu}})e^{-i(\gamma-\beta)} 
& c_{\nu23}c_{\nu13}\\ 
\end{array}
\right) \ .\nonumber\\
& &
\label{majorana}
\end{eqnarray}
Here, \(\alpha_i^e\) comes from the rephasing in the charged lepton fields 
to make the choice of phase convention.
The CP-violating phase \(\delta_{\nu}\), the additional Majorana 
phase $\beta$ and $\gamma$ \cite{bilenky,Doi} 
in the representation Eq.~(\ref{majorana}) are calculable and obtained as
\begin{eqnarray}
\delta_\nu  &=& \mbox{arg}
          \left[
             \frac{U_{12}U_{22}^*}{U_{13}U_{23}^*} + 
             \frac{|U_{12}|^2}{1-|U_{13}|^2}
          \right] 
 \simeq \pi ,\nonumber \\ 
\beta & =& \mbox{arg} \left( \frac{U_{12}}{U_{11}}\right) 
  \simeq -\pi/2, \\
 \gamma  & =& \mbox{arg} \left( \frac{U_{13}}{U_{11}}e^{i\delta_\nu}\right) 
  \simeq \pi/2\ ,\nonumber
\end{eqnarray}
by using the relation \(m_e \ll m_\mu \ll m_\tau \).

\par
We also predict the averaged neutrino mass 
$\langle m_\nu \rangle$ which appears in the neutrinoless 
double beta decay~\cite{Doi} as follows:
\begin{eqnarray}
\langle m_\nu \rangle  &\equiv&  \left| m_1 U_{11}^2 +m_2 U_{12}^2 +m_3 U_{13}^2 \right| =\frac{m_e m_3}{2m_\mu} \nonumber\\
&=& (0.89-1.4)\times10^{-4}\mbox{ eV}.
\end{eqnarray}
This value of $\langle m_\nu \rangle$ is too small to be observed in near future experiments~\cite{Ejiri}.
\section{summary}
\par 
We have investigated a Hermite mass matrix model given in Eqs.~(2.1)--(2.6), 
in which the mass matrices for quarks and charged leptons are assumed to have 
a term in which the $2 \leftrightarrow 3$ symmetry is maximally broken. 
The mass matrices for up quarks, down quarks, charged leptons, and neutrinos 
have a common structure as shown by $\widehat{M_f}$ in Eq.~(2.7) when it is expressed 
after  rebasing of the quark and lepton fields. 
The large lepton mixing angle between the second and third generation is realized with no fine tuning in our model.
The model is almost consistent  with the present data in the quark as well as lepton sectors.
The model also predicts  $|U_{13}|^2 \simeq \frac{m_e}{2m_\mu}=2.4\times10^{-3}$ for the lepton mixing matrix element $U_{13}$, and  
neutrino masses shown in Eq.~(4.26) are obtained from the neutrino oscillation data 
for \(\theta_{\mbox{{\tiny sol}}}\),  \(\Delta m_{23}^2\), and \(\Delta m_{12}^2\).
We also predict $\langle m_\nu \rangle=(0.89-1.4)\times10^{-4}$ eV 
for the averaged neutrino mass which appears in the neutrinoless double beta decay.


\begin{acknowledgments}
This work of K.M. was supported by the JSPS, No. 15-3700.
\end{acknowledgments}
\appendix

\section{diagonalization of mass matrix $\widehat{M_f}$}
For the purpose of making this paper self-contained, here we summarize the 
diagonalization of mass matrix $\widehat{M_f}$ (f=u,d,e and $\nu$) defined by
\begin{equation}
\widehat{M_f}
=\left(
\begin{array}{lll}
\ 0 & \ A_f & \ 0 \\
\ A_f & \ B_f & \ C_f \\
\ 0 & \ C_f & \ D_f \\
\end{array}
\right),
\end{equation}
for up quarks, down quarks, charged leptons, and neutrinos.
\par
\subsection{mass matrix $\widehat{M_f}$ for quarks and charged leptons}
For quarks and charged leptons (f=u, d, and e), let us take 
a following choice for $\widehat{M_f}$: 
\begin{equation}
\widehat{M_f}=
\left(
\begin{array}{lll}
\ 0 & \ A_f & \ 0 \\
\ A_f & \ B_f & \ C_f \\
\ 0 & \ C_f & \ D_f \\
\end{array}
\right)
=
\left(
        \begin{array}{ccc}
        0&\scriptstyle{\sqrt{\frac{m_{1f}m_{2f}m_{3f}}{m_{3f}-m_{1f}}}}&0\\
        \scriptstyle{\sqrt{\frac{m_{1f}m_{2f}m_{3f}}{m_{3f}-m_{1f}}}}
        & m_{2f} & 
    \scriptstyle{\sqrt{\frac{m_{1f}m_{3f}(m_{3f}-m_{2f}-m_{1f})}{m_{3f}-m_{1f}}}}\\
        0& \scriptstyle{\sqrt{\frac{m_{1f}m_{3f}(m_{3f}-m_{2f}-m_{1f})}{m_{3f}-m_{1f}}}}& 
    m_{3f}-m_{1f}
        \end{array}
\right) . 
\end{equation}
This is diagonalized  by an orthogonal matrix $O_{f}$ as (see Ref\cite{Koide}\cite{Matsuda})
\begin{equation}
O_{f}^T\left(
\begin{array}{lll}
\ 0 & \ A_{f} & \ 0 \\
\ A_{f} & \ B_{f} & \ C_{f} \\
\ 0 & \ C_{f} & \ D_{f} \\
\end{array}
\right)
O_{f}
=\left(
\begin{array}{lll}
-m_{1f} & \  & \  \\
\  & m_{2f} & \  \\
\  & \  & m_{3f} \\
\end{array}
\right).
\end{equation}
Here $m_{if} (i=1,2,3)$ are eigenmasses and \(O_{f}\) is given by 
\begin{eqnarray}
O_f
&=&
\left(
\begin{array}{ccc}
{\sqrt{\frac{m_{2f}m_{3f}^2}{(m_{2f}+m_{1f})(m_{3f}^2-m_{1f}^2)}}}&
{\sqrt{\frac{m_{1f}m_{3f}(m_{3f}-m_{2f}-m_{1f})}{(m_{2f}+m_{1f})(m_{3f}-m_{2f})(m_{3f}-m_{1f})}}}&
{\sqrt{\frac{m_{1f}^2m_{2f}}{(m_{3f}-m_{2f})(m_{3f}^2-m_{1f}^2)}}} \\
-{\sqrt{\frac{m_{1f}m_{3f}}{(m_{2f}+m_{1f})(m_{3f}+m_{1f})}}}&
{\sqrt{\frac{m_{2f}(m_{3f}-m_{2f}-m_{1f})}{(m_{2f}+m_{1f})(m_{3f}-m_{2f})}}}&
{\sqrt{\frac{m_{1f}m_{3f}}{(m_{3f}-m_{2f})(m_{3f}+m_{1f})}}} \\
{\sqrt{\frac{{{m_{1f}}^2}(m_{3f}-m_{2f}-m_{1f})}{(m_{2f}+m_{1f})(m_{3f}^2-m_{2f}^2)}}}&
-{\sqrt{\frac{m_{1f}m_{2f}m_{3f}}{(m_{3f}-m_{2f})(m_{2f}+m_{1f})(m_{3f}-m_{1f})}}}&
{\sqrt{\frac{(m_{3f})^2(m_{3f}-m_{2f}-m_{1f})}{(m_{3f}^2-m_{2f}^2)(m_{3f}-m_{2f})}}} 
\end{array}
\right) \nonumber \\
&\simeq&
\left(
        \begin{array}{ccc}
        1& \displaystyle{\sqrt{\frac{m_{1f}}{m_{2f}}}}&
         \displaystyle{\sqrt{\frac{m_{1f}m_{2f}^2}{m_{3f}^3}}}\\
        \displaystyle{-\sqrt{\frac{m_{1f}}{m_{2f}}}}
        & 1 & \displaystyle{\sqrt{\frac{m_{1f}}{m_{3f}}}}\\
        \displaystyle{\sqrt{\frac{{\scriptstyle{m_{1f}^2}}}{m_{2f}m_{3f}}}}
        & \displaystyle{-\sqrt{\frac{m_{1f}}{m_{3f}}}} & 1
        \end{array}
\right) \qquad (\mbox{for }m_{3f} \gg m_{2f} \gg m_{1f}).
\label{eq990114}
\end{eqnarray}
Here $m_{iu}$, $m_{id}$, and $m_{ie}\ (i=1,2,3) $ are, respectively, masses of up quarks, down quarks, charged leptons, and neutrinos, 
which we shall denoted as $(m_u, m_c,m_t)$, $(m_d, m_s,m_b)$, and $(m_e, m_\mu,m_\tau)$.
\par
\subsection{mass matrix $\widehat{M_\nu}$ for neutrinos}
For neutrinos (f=$\nu$) we choose : 
\begin{equation}
\widehat{M_{\nu}}=
\left(
\begin{array}{lll}
\ 0 & \ A_{\nu} & \ 0 \\
\ A_{\nu} & \ B_{\nu} & \ 0 \\
\ 0 & \ 0 & \ D_{\nu} \\
\end{array}
\right)
=
\left(
        \begin{array}{ccc}
        0&\sqrt{m_1m_2}&0\\
        \sqrt{m_1m_2} & m_2-m_1 & 0\\
        0& 0& m_3
        \end{array}
\right). 
\end{equation}
Note we take $C_{\nu}=0$.
This $\widehat{M_{\nu}}$ is diagonalized as 
\begin{equation}
O_{\nu}^T\left(
\begin{array}{lll}
\ 0 & \ A_{\nu} & \ 0 \\
\ A_{\nu} & \ B_{\nu} & \ 0 \\
\ 0 & \ 0 & \ D_{\nu} \\
\end{array}
\right)
O_{\nu}
=\left(
\begin{array}{lll}
-m_{1} & \  & \  \\
\  & m_{2} & \  \\
\  & \  & m_{3} \\
\end{array}
\right),
\end{equation}
where $m_{i} (i=1,2,3)$ are neutrino masses and the orthogonal matrix \(O_{\nu}\) is given by 
\begin{equation}
O_{\nu}
=
\left(
        \begin{array}{ccc}
        \sqrt{\frac{m_2}{m_2+m_1}}& \sqrt{\frac{m_1}{m_2+m_1}}&
         0\\
        -\sqrt{\frac{m_1}{m_2+m_1}}
        & \sqrt{\frac{m_2}{m_2+m_1}} & 0\\
        0
        & 0& 1
        \end{array}
\right).
\end{equation}
\section{Evolution effect}
We have estimated the evolution effects for the CKM matrix elements from  the electroweak scale \(\mu=m_Z\) to the unification scale \(\mu=M_X\)
by using the two-loop renormalization group equation (RGE) 
[minimal supersymmetric standard model with tan$\beta$=10 case] for the Yukawa coupling constants.   
In the numerical calculations, we have used the following running quark masses at \(\mu=m_Z\)  
and at \(\mu=M_X\) \cite{Fusaoka}:
\begin{equation}
\begin{array}{ll}
m_u(m_Z)=2.33^{+0.42}_{-0.45}\, \mbox{MeV},& 
m_d(m_Z)=4.69^{+0.60}_{-0.66}\, \mbox{MeV},  \\
m_c(m_Z)=677^{+56}_{-61}\, \mbox{MeV},  &
m_s(m_Z)=93.4^{+11.8}_{-13.0}\, \mbox{MeV}, \\ 
m_t(m_Z)=181\pm13\, \mbox{GeV}, &
m_b(m_Z)=3.00\pm0.11\, \mbox{GeV}. 
\end{array}
\end{equation}
\begin{equation}
\begin{array}{ll}
m_u(M_X)=1.04^{+0.19}_{-0.20}\, \mbox{MeV},& 
m_d(M_X)=1.33^{+0.17}_{-0.19}\, \mbox{MeV}, \\
m_c(M_X)=302^{+25}_{-27}\, \mbox{MeV}, &
m_s(M_X)=26.5^{+3.3}_{-3.7}\, \mbox{MeV}, \\ 
m_t(M_X)=129^{+196}_{-40}\,  \mbox{GeV}, &
m_b(M_X)=1.00\pm0.04\, \mbox{GeV}. 
\end{array}
\end{equation}
\par
We have calculated numerical values of the CKM mixing matrix elements at \(\mu=M_X\) 
from their observed values at \(\mu=m_Z\). 
Namely using as inputs the observed quark mixing angles and the $CP$ violating phase at \(\mu=m_Z\) 
given by 
\begin{eqnarray}
\sin\theta_{12}(m_Z) &=& 0.2243 \pm 0.0016, \quad \sin\theta_{23}(m_Z) = 0.0413 \pm 0.0015, \nonumber\\
\sin\theta_{13}(m_Z) &=& 0.0037 \pm 0.0005, \quad \delta(m_Z) = 60^\circ \pm 14^\circ, \label{CKM_mz}
\end{eqnarray}
we obtain the following numerical values for the mixing angles 
and the magnitude of the mixing matrix elements at \(\mu=M_X\) \cite{Matsuda3}:
\begin{eqnarray}
\sin\theta_{12}^0 &=& 0.2226 - 0.2259, \quad 
\sin\theta_{23}^0 = 0.0295 - 0.0383, \nonumber \\
\sin\theta_{13}^0 &=& 0.0024 - 0.0038, \quad 
\delta^0    \  = \ 46^\circ   - 74^\circ , \label{eq1210-01}\\
|V^0| &=& 
\left(
\begin{array}{ccc}
0.9741-0.9749 & 0.2226-0.2259 & 0.0024-0.0038 \\
0.2225-0.2259 & 0.9734-0.9745 & 0.0295-0.0387 \\
0.0048-0.0084 & 0.0289-0.0379 & 0.9993-0.9996
\end{array}
\right). \label{eq1118-01} 
\end{eqnarray}



\end{document}